\documentclass[pra,showpacs,aps,twocolumn,superscriptaddress]{revtex4-1}

\usepackage{amsmath, amssymb}
\usepackage{graphicx}
\graphicspath{{./figures/}{./figuretmp/}}

\newcommand{\tr}{\mathop{\mathrm{tr}}\nolimits}
\begin{document}
\title{Full-counting statistics of time-dependent conductors}

\author{M\'onica Benito}
\affiliation{Instituto de Ciencia de Materiales de Madrid, CSIC, 28049
Madrid, Spain}
\author{Michael Niklas}
\affiliation{Institut f\"ur Theoretische Physik, Universit\"at Regensburg,
93040 Regensburg, Germany}
\author{Sigmund Kohler}
\affiliation{Instituto de Ciencia de Materiales de Madrid, CSIC, 28049
Madrid, Spain}

\date{\today}


\begin{abstract}
We develop a scheme for the computation of the full-counting statistics of
transport described by Markovian master equations with an arbitrary time
dependence.  It is based on a hierarchy of generalized density operators,
where the trace of each operator yields one cumulant.  This direct relation
offers a better numerical efficiency than the equivalent number-resolved
master equation.  The proposed method is particularly useful for conductors
with an elaborate time-dependence stemming, e.g., from pulses or
combinations of slow and fast parameter switching.  As a test bench for the
evaluation of the numerical stability, we consider time-independent problems
for which the full-counting statistics can be computed by other means.  As
applications, we study cumulants of higher order for two time-dependent
transport problems of recent interest, namely steady-state coherent
transfer by adiabatic passage (CTAP) and
Landau-Zener-St\"uckelberg-Majorana (LZSM) interference in an open double
quantum dot.
\end{abstract}

\maketitle

\section{Introduction}

Current fluctuations, while probably disturbing in technical applications,
can be useful for understanding quantum mechanical transport processes
\cite{Beenakker2003a}.  For instance, an open transport channel with
transmission close to unity leads to sub Poissonian noise, while super
Poissonian noise may hint on bistabilities \cite{Blanter2000a}.
External driving fields enable the control of the noise level via
the driving amplitude and frequency \cite{Camalet2003a}.
Particular examples of such driven conductors with low current noise are
pumps that transport a fixed charge per cycle \cite{Fricke2013a,
Kaestner2015a, Croy2016a}.  Moreover, noise measurements may provide
evidence for the correct operation of protocols that induce a steady-state
version \cite{Huneke2013a} of coherent transport by adiabatic passage
\cite{Greentree2004a, Eckert2004a, MenchonEnrich2016a}.

Current fluctuations can be characterized by the low-frequency
limit of the current correlation function which corresponds to the variance
of the transported charge \cite{MacDonald1949a}.  This allows one to introduce
the Fano factor as a dimensionless measure for the noise level using the
Poisson process as reference \cite{Blanter2000a}.  Going beyond the
variance, one may consider the full counting statistics
of the transported electrons \cite{Buttiker1992a, Levitov1993a,
Blanter2000a, Bagrets2003a, Belzig2005a} or the related waiting-time
distribution of consecutive transport events \cite{Brandes2008a}.

For master equation descriptions of \textit{time-independent} transport,
the calculation of the full counting statistics can be formulated as a
non-Hermitian eigenvalue problem with a subsequent computation
of derivatives with respect to a counting variable \cite{Bagrets2003a}.
For systems with very few degrees of freedom, this may provide all
cumulants analytically \cite{Bagrets2003a, Belzig2005a}.  For a numerical
treatment, however, one likes to avoid the probably unstable computation
of higher-order derivatives, which can be achieved by an iterative scheme
based on Rayleigh-Schr\"odinger perturbation theory \cite{Flindt2008a,
Flindt2010a}.

These eigenvalue based methods are generally not applicable for conductors
with an arbitrary \textit{time-dependence}, so that one has to seek for
alternatives.  One option is a number-resolved master equation in which the
number of transported electrons is introduced as an additional degree of
freedom \cite{Gurvitz1996a, Kubala2016a, Cerrillo2016a}.  However, the distribution of
this number may be rather broad and, thus, the computational effort may
become tremendous.  A more efficient approach is based on a density
operator like object that contains information about the second moment of
the transported charge \cite{Sanchez2008c}.  A numerical solution of the
corresponding equations of motion provides the current and its variance
with moderate numerical effort.  With the present work we generalize this
idea and derive a propagation method for computing current cumulants up to
a given order.

Our paper is structured as follows. In Sec.~\ref{sec:formalism}, we
introduce a master equation description of the full-counting
statistics and derive our iteration scheme.  In Sec.~\ref{sec:testcases},
we explore the numerical stability of our method for two time-independent
test cases and finally in Sec.~\ref{sec:applications} study cumulants of
higher order for two driven models of recent interest, namely steady-state
CTAP and LZSM interference.

\section{Generalized master equation}
\label{sec:formalism}

We consider transport problems that can be captured by a master equation of
the form
\begin{equation}
\label{meq}
\dot\rho = -\frac{i}{\hbar}[H(t),\rho] + \sum_\ell\mathcal{L}_\ell(t)\rho
\equiv \mathcal{L}(t)\rho ,
\end{equation}
where $H(t)$ accounts for the coherent quantum dynamics of a central
conductor such as a quantum dot array driven by time-dependent gate
voltages.  The conductor is coupled to two or more electron reservoirs that
allow for incoherent electron tunneling from and to the reservoirs.  These
processes are described by the generally also time-dependent super operators
$\mathcal{L}_\ell$ which contain the forward and backward current super
operators $\mathcal{J}_\ell^+$ and $\mathcal{J}_\ell^-$, respectively.
For a specific example of these super operators, see
Sec.~\ref{sec:testcases}.

\subsection{Counting variable}

The electron transport can be considered as a stochastic process with the
random variable $N_\ell$, the net number of electrons transported to
lead $\ell$ or, equivalently, the electron number in that lead (to achieve
a compact notation, we henceforth suppress the lead index $\ell$ and the
time argument).  Its statistical properties can be captured by the moment
generating function
\begin{equation}
\label{Z}
Z(\chi) = \langle e^{i\chi}\rangle
= \sum_{k=0}^\infty \frac{(i\chi)^k}{k!} \mu_k ,
\end{equation}
with the moments $\mu_k = \langle N^k\rangle = (\partial/\partial i\chi)^k
Z|_{\chi=0}$, while their irreducible parts, the cumulants $\kappa_k$, are
generated from $\log Z(\chi)$ \cite{vanKampen1992a}.  For Markovian
time-independent transport problems, the cumulants eventually grow linearly
in time \cite{Bagrets2003a} which motives the definition of the \textit{current
cumulants} as the time derivatives $c_k = \dot\kappa_k$, which are our main
quantities of interest.  Their generating function reads
\begin{equation}
\label{phi}
\phi(\chi) = \frac{d}{dt}\log Z(\chi)
= \sum_{k=1}^\infty \frac{(i\chi)^k}{k!} c_k ,
\end{equation}
which implies $c_k = (\partial/\partial i\chi)^k \phi|_{\chi=0}$.

While the master equation \eqref{meq} contains the full information about
the central conductor, the leads degrees of freedom have been traced out in
course of its derivation.  To nevertheless keep track of the electron
number in lead $\ell$, one multiplies the full density
operator by a counting factor $e^{i\chi N}$ for the lead electrons
to obtain the generalized density operator $R(\chi)$.  It relates to the
moment generating function \eqref{Z} via $\tr R(\chi) = Z(\chi)$ and obeys
the generalized master equation \cite{Bagrets2003a}
\begin{equation}
\label{gme}
\dot{R}(\chi) = [\mathcal{L} + \mathcal{J}(\chi)] R(\chi) .
\end{equation}
The additional term
\begin{equation}
\mathcal{J}(\chi) = (e^{i\chi}-1)\mathcal{J}^+ +(e^{-i\chi}-1)\mathcal{J}^-
\end{equation}
is composed of the forward and the backward current operators
$\mathcal{J}^\pm$ mentioned above.

\subsection{Hierarchy of master equations}

The generalized master equation \eqref{gme} together with the generating
functions \eqref{Z} and \eqref{phi} in principle already provides the
current cumulants $c_k$.  The direct numerical evaluation of these
expressions, however, is hindered by two obstacles.  First, the numerical
computation of derivatives becomes increasingly difficult with the
order.  Second, the relation between cumulants and moments is known only
implicitly via the Taylor series for $Z(\chi)$ and $\phi(\chi)$.  Therefore
we have to bring the generalized master equation to a form that is more
suitable for extracting information about the $c_k$.

We start by writing the current cumulant generating function in terms of
the generalized density operator $R(\chi)$.  From the definitions $\phi =
\log\dot Z$ and $Z=\tr R(\chi)$ together with the generalized master
equation \eqref{gme} follows straightforwardly
\begin{equation}
\phi(\chi) = \frac{1}{Z(\chi)} \tr\mathcal{J}(\chi) R(\chi)
= \tr\mathcal{J}(\chi) X(\chi)
\end{equation}
(notice that $\tr\mathcal{L}\ldots = 0$) with the auxiliary operator
\begin{equation}
X(\chi) = \frac{1}{Z(\chi)} R(\chi) .
\end{equation}
Moreover, we find the equation of motion
\begin{equation}
\label{dotXchi}
\dot X(\chi) = \mathcal{L}X(\chi) + [\mathcal{J}(\chi) - \phi(\chi)] X(\chi) .
\end{equation}
We continue by substituting the dependence on the continuous counting
variable $\chi$ by the Taylor coefficients $X_k$ and $\mathcal{J}_k$ which we
define via the series
$X(\chi) = \sum_{k=0}^\infty (i\chi)^k X_k/k!$ and
$\mathcal{J}(\chi) = \sum_{k=1}^\infty (i\chi)^k \mathcal{J}_k/k!$.
Notice that $\mathcal{J}(0)=0$ such that $\mathcal{J}_0=0$ while for $k>0$,
$\mathcal{J}_k = \mathcal{J}^++(-1)^k\mathcal{J}^-$.
Finally, we obtain from Eqs.~\eqref{phi} and \eqref{dotXchi} the hierarchy of
equations
\begin{align}
\label{iteration1}
c_k ={}& \sum_{k'=0}^{k-1}\binom{k}{k'} \tr\mathcal{J}_{k-k'}X_{k'} ,
\\
\label{iteration2}
\dot X_k ={}& \mathcal{L}X_k +\sum_{k'=0}^{k-1}\binom{k}{k'}
(\mathcal{J}_{k-k'} - c_{k-k'})X_{k'} .
\end{align}
It constitutes the central formal achievement of this paper and forms the
basis of the numerical results presented below.

Two features are worth being emphasized. First, in the limit
$\chi\to 0$, $X(\chi)$ becomes the reduced density operator, i.e., for
$k=0$, Eq.~\eqref{iteration2} is identical to the master equation
\eqref{meq}.  Second, as an important consequence of $\mathcal{J}_0=0$ and
$c_0=0$, the summations on the r.h.s.\ of these equations terminate at
$k'=k-1$, which implies that $X_{k}$ and $c_k$ depend only on terms of lower
order.  This enables the truncation at arbitrary order and, thus, the iterative
computation of the current cumulants.

The numerical effort of our scheme can be estimated as follows. Let us
assume that (if necessary after a full or a partial \cite{Darau2009a}
rotating-wave approximation) the Liouvillian $\mathcal{L}$ can be written
as a $d\times d$-matrix and that its smallest decay rate is
$\gamma_\text{min}$.  Then to compute the first $k_\text{max}$ cumulants,
we have to propagate $k_\text{max}d$ scalar equations for a time
$\tau\approx 3/\gamma_\text{min}$, where one is typically interested in the
first $k_\text{max}=5$--$10$ cumulants.

To highlight the efficiency of our method, we compare this effort with that
of the number-resolved master equation \cite{Gurvitz1996a, Kubala2016a,
Cerrillo2016a}, for which the
density operator is extended by a variable $n = 0,\ldots,n_\text{max}$ that
accounts for the number of transported electrons.  In the Markovian case,
coherences between different $n$ do not play a role, such that one
essentially has to replace $\rho$ by the $n_\text{max}+1$ density operators
$\rho^{(n)}$, where $\tr\rho^{(n)}$ is the probability that $n$ electrons
have arrived at a certain lead.  During the time $\tau$, on average $I\tau$
electrons flow, so that one would have to employ a number-resolved master
equation with $n_\text{max}\approx 2I\tau = 6I/\gamma_\text{min}$, i.e.,
one has to integrate $\sim 6Id/\gamma_\text{min}$ scalar equations.  This
means that whenever $I\gtrsim\gamma_\text{min}$, our method outperforms this
alternative significantly.  This is for example the case when the system
infrequently switches between two states with different conductance
\cite{Belzig2005a, Koch2005a, Lambert2015a}.  A further advantage of our
method is that it provides direct access to the cumulants, such that the
detour via the moments can be avoided.

\subsection{Relation to the iterative scheme for time-independent transport}

Equations~\eqref{iteration1} and \eqref{iteration2} resemble the iterative
scheme derived in Refs.~\cite{Flindt2008a, Flindt2010a} for the cumulants
of \textit{time-independent} transport problems.  Let us therefore
establish a connection between both methods.
If $\mathcal{L}$ is time-independent, the original master equation
possesses a stationary solution $\rho_\infty$ which for $k=0$ also solves
Eq.~\eqref{iteration2}.  For $k>0$, we make use of the fact that $\tr
X(\chi)=1$ which implies $\tr X_k=\delta_{k,0}$.  Consequently,
Eq.~\eqref{iteration2} possesses also for $k>0$ a stationary solution.
Formally it can be written with the help of the pseudo-inverse of the
Liouvillian $\mathcal{Q}/\mathcal{L}$, where $\mathcal{Q} =
\mathbf{1}-\rho_\infty\tr$ projects to the subspace in which $\mathcal{L}$
is regular.  Therefore, the condition $\dot X_k=0$ together with $\tr X_k =
\delta_{k,0}$ results in
\begin{equation}
X_k = \frac{\mathcal{Q}}{\mathcal{L}} \sum_{k'=0}^{k-1}\binom{k}{k'}
(\mathcal{J}_{k-k'} - c_{k-k'})X_{k'} ,
\label{iteration2_stat}
\end{equation}
while $X_0=\rho_\infty$.  Equations~\eqref{iteration1} and
\eqref{iteration2_stat} represent the time-independent Markovian limit of
the known iteration scheme for cumulants \cite{Flindt2008a, Flindt2010a}.

\section{Time-independent models as test cases}
\label{sec:testcases}

Before addressing time-dependent transport problems, let us start with two
time-independent systems which can be solved either analytically or with
the iteration scheme of Ref.~\cite{Flindt2008a}.  This allows us to draw
conclusions about the numerical stability of our method.
To this end, we consider the cumulant ratio
\begin{equation}
F_k = c_{k+1}/c_k,
\end{equation}
where $F_1$ is the Fano factor.

Despite the general validity of our formalism, in all applications, we
consider an array of $n$ quantum dots with the first dot coupled to an
electron source $S$, while the last site is coupled to a drain $D$.  Then
the dot-lead tunnelings can be written as $\mathcal{L}_\text{dot-lead} =
\Gamma_S\mathcal{D}(c_1^\dagger) + \Gamma_D\mathcal{D}(c_n)$ with the
Lindblad form $\mathcal{D}(x)\rho = x\rho x^\dagger - \frac{1}{2}x^\dagger
x\rho -\frac{1}{2}\rho x^\dagger x$ and the tunnel rates $\Gamma_{S/D}$.
We evaluate the current at the source, $\ell=S$, such that the forward
current operator becomes $\mathcal{J}_S^+\rho \equiv \mathcal{J}_S^+ =
\Gamma_S c_1^\dagger\rho c_1$, while the backward current operator
$\mathcal{J}_S^-$ vanishes.

\subsubsection{Single-electron transistor}

\begin{figure}
\includegraphics[width=.9\columnwidth]{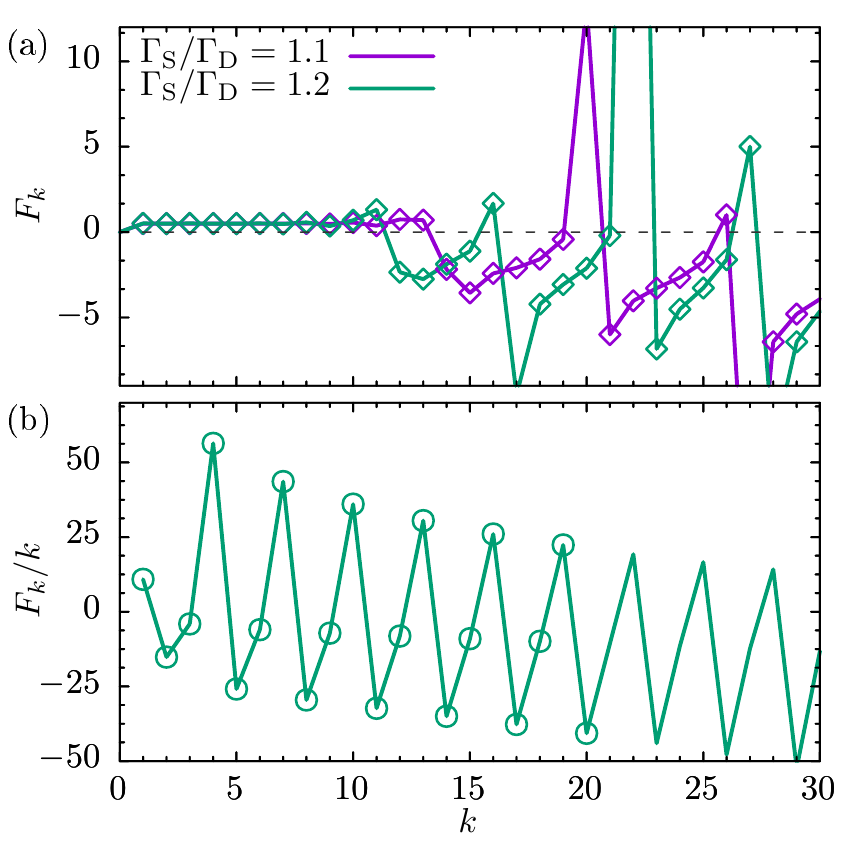}
\caption{Cumulant ratios $F_k=c_{k+1}/c_k$ for time-independent test cases.
The symbols are obtained with our propagation method, while
the lines interpolate the results of the iteration scheme based on
Eq.~\eqref{iteration2_stat}.
(a) Asymmetric single-electron transistor for large bias and various
dot-lead rates $\Gamma_{S/D}$.
(b) Triple quantum dot in ring configuration with $\Gamma_S =\Gamma_D
=0.1\Delta$, where dot~2 is detuned by $\epsilon=10\Delta$.
For graphical reasons, we plot $F_k/k$.
}
\label{fig:test}
\end{figure}

One of the simplest transport setups is the single-electron transistor
which consists of a resonant level between two strongly biased
leads.  It can be occupied by at most one electron so that
the Liouvillian and the forward current operator read
\begin{equation}
\mathcal{L} =
\begin{pmatrix} -\Gamma_S & \Gamma_D\\ \Gamma_S & -\Gamma_D \end{pmatrix} ,
\qquad
\mathcal{J}^+ =
\begin{pmatrix} 0 & 0 \\ \Gamma_S & 0 \end{pmatrix} ,
\end{equation}
respectively.
For the symmetric case, $\Gamma_S=\Gamma_D\equiv\Gamma$, the cumulants of
the single-electron transistor are known analytically as $c_k =
2^{-k}\Gamma$ \cite{Bagrets2003a}, which makes this system an ideal test
case.  Consequently, all cumulant ratios $F_k = 1/2$ are identical to the
Fano factor.  For any $\Gamma_S \neq \Gamma_D$, the cumulants cannot be
written in a closed form, but exhibit a generic behavior:  While cumulants
of low order reflect the nature of the transport process, high-order
cumulants oscillate in a universal manner \cite{Flindt2009a}.  Therefore
the symmetric case with its constant $F_k=1/2$ is rather special and should
be sensitive to numerical errors.

By solving Eqs.~\eqref{iteration1} and \eqref{iteration2} numerically, we
have found that for $\Gamma_S=\Gamma_D\equiv\Gamma$, the first $\gtrsim 30$
cumulant ratios agree with the analytical prediction with a
precision $\lesssim 1\%$ (not shown).  For slight asymmetries, we
compare in Fig.~\ref{fig:test}(a) our results with those obtained by the
traditional iteration scheme.  Both agree rather well also for orders at
which the cumulants exhibit universal oscillations.

\subsubsection{Triple quantum dot in a ring configuration}

As a further test case, we consider a ring of three quantum dots, where
dots~1 and 3 are is coupled to source and drain, respectively,
while dot~2 is detuned by a onsite energy $\epsilon$.  The
corresponding single-particle Hamiltonian reads
\begin{equation}
H = \begin{pmatrix} 0 & \Delta & \Delta \\ \Delta & \epsilon & \Delta \\
\Delta & \Delta & 0 \end{pmatrix} .
\end{equation}
Transport may thus occur by direct tunneling from the first to the last dot
or via dot~2.  For strong detuning, $\epsilon\gg\Delta$, the latter path
has the effective tunnel matrix element $\Delta^2/\epsilon \ll \Delta$.
Thus in the limit of strong Coulomb repulsion, the situation is that of a
slow and a fast channel which block one another.  This typically leads to
bunching visible in a super Poissonian Fano factor \cite{Belzig2005a}.  The
triple quantum dot ring combines several difficulties such as different
time scales, quantum interference, and cumulants that grow exponentially
with their index \cite{Dominguez2010a}.  The corresponding stiff
differential equations represent challenging test cases for propagation
methods.

In Fig.~\ref{fig:test}(b) we again compare the results of our method with
those of the iteration of Eq.~\eqref{iteration2_stat}.  As for the double
quantum dot, we find that for the first 20 cumulants, the results of both
methods are practically indistinguishable.  Owing to the mentioned
difficulties, however, calculations for more than roughly 15
cumulants require a rather high numerical precision and, thus, are
time consuming.  Nevertheless, we can conclude that for the experimentally
relevant orders, our scheme is still efficient and numerically stable.

\section{Applications}
\label{sec:applications}

To demonstrate the practical use of our time-dependent iteration scheme, we
apply it to various physical situations that have been studied recently,
i.e., we generalize previous calculations of the current or the Fano
factor to cumulants of higher order.

\subsection{Steady-state coherent transfer by adiabatic passage}

\begin{figure}
\includegraphics[width=.9\columnwidth]{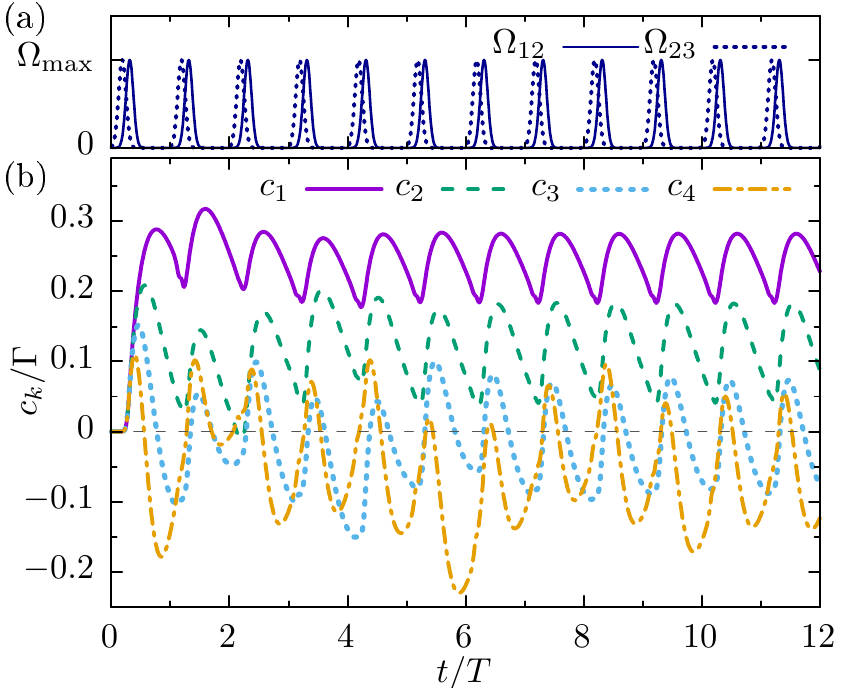}
\caption{(a) Pulsed tunnel matrix elements defined in Eq.~\eqref{OmegaCTAP}
which lead to an adiabatic passage of electrons from dot 1 to dot 3.  Each
pulse has a width $\sigma=T/16$.  The delay within a double pulse is
$\Delta t=T/8$, while the time between the pairs is $T=40/\Omega_\text{max}$.
(b) Corresponding time evolution of the current cumulants $c_k$,
$k=1,\ldots,4$, for the dot-lead rates $\Gamma_S =\Gamma_D
=0.05\Omega_\text{max}$ \cite{onHuneke}.
}
\label{fig:ctap}
\end{figure}
\begin{figure}
\includegraphics[width=.9\columnwidth]{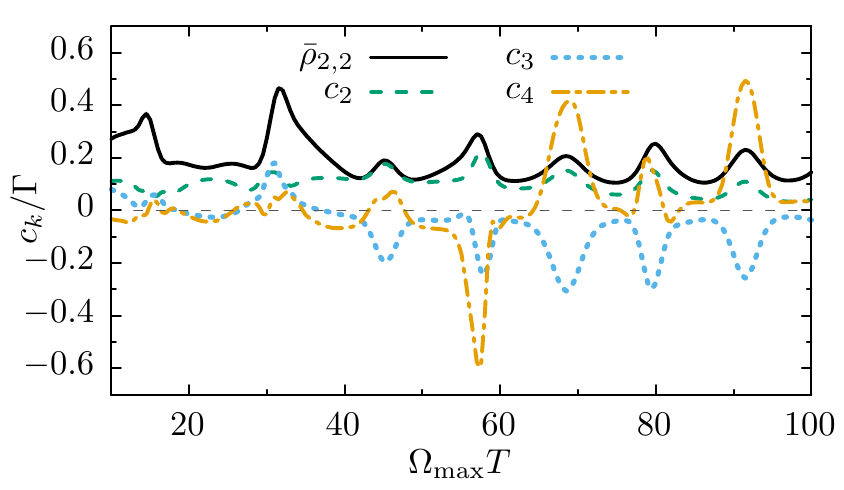}
\caption{Time-averaged population of the central dot for steady-state CTAP
as a function of the driving period $T$ together with the cumulants 
$c_{k}$ for $k=2,3,4$.  All other parameters are as in Fig.~\ref{fig:ctap}.
}
\label{fig:ctapF}
\end{figure}

Let us consider a triple quantum dot described by the single-particle
Hamiltonian
\begin{equation}
H(t) =
\begin{pmatrix} 0 & \Omega_{12}(t) & 0 \\
\Omega_{12}(t) & 0 & \Omega_{23}(t) \\
0 & \Omega_{23}(t) & 0 \end{pmatrix}.
\label{Hctap}
\end{equation}
If the tunnel couplings $\Omega_{ij}$ are switched adiabatically slowly,
the system may follow the adiabatic eigenstate $\propto
(\Omega_{23},0,-\Omega_{12})^T$.  In this way, it is possible to
transfer
an electron from the first dot to the last dot without populating the
middle dot \cite{Greentree2004a}, an effect known as CTAP.  This non-local
version of an optical Lambda transition \cite{Vitanov2001a} has also been
predicted for atoms in multi-stable traps \cite{Eckert2004a,
MenchonEnrich2016a}.

Experimental evidence of the direct tunneling from the first to the last
dot is hindered by the backaction of a population measurement, which
creates decoherence \cite{Rech2011a} and, thus, may induce the effect that
one wishes to demonstrate.  To circumvent this problem, is has been
suggested \cite{Huneke2013a} to contact the triple quantum dot to an
electron source and drain and to employ the sequence of double Gauss pulses
\begin{equation}
\label{OmegaCTAP}
\Omega_{12/23}(t) = \sum_{n=0}^\infty \Omega_\text{max} \exp\Big[
-\frac{(t\mp\Delta t/2 -nT)^2}{2\sigma^2}\Big]
\end{equation}
with width $\sigma$, delay $\Delta t$, and repetition time $T$, as is
sketched in Fig.~\ref{fig:ctap}(a).  Notice the so-called counter intuitive
order of the pulses in which the tunnel matrix element $\Omega_{23}$ is
active before $\Omega_{12}$.  In the ideal case, this sequence will
lead to the transport of one electron per double pulse and, thus, induce a
current with a low Fano factor which may serve as experimental verification
of CTAP.

While in Ref.~\cite{Huneke2013a} only the second current cumulant have been
considered, we here focus on cumulants of higher order.  We again assume
that Coulomb repulsion inhibits the occupation with more than one electron.
Then we have to add the empty state to the Hamiltonian \eqref{Hctap}, while
the dissipative parts of the Liouville equation and the current operator
remain the same as in the last section.

Figure~\ref{fig:ctap}(b) shows the time evolution of the first four current
cumulants.  After a transient stage of roughly $10T$, the dynamics assumes
its long time limit, from which we compute the steady state values of the
cumulants as the average over the driving period.  The time evolution
illustrates that generally the duration of the transient stage increases
with the cumulant order.

The central issue of verifying CTAP via noise measurements is the
correlation between the Fano factor and the population of the middle dot as
a function of the driving period $T$.  By contrast, the current correlates
only weakly with the population and cannot serve as indicator
\cite{Huneke2013a}.  Notice that a non-trivial value for the correlation
coefficient requires a non-monotonic variation of both curves, which indeed
is the case.  Going beyond this, we plot in Fig.~\ref{fig:ctapF} the
corresponding cumulants of higher order averaged over one driving period in
the long-time limit.  We find that the third cumulant also correlates with
the occupation, while for the fourth cumulant only the absolute value
behaves in this way.  Interestingly enough, the profile of $c_3$ and $c_4$
cumulant is even sharper than that of the zero-frequency noise $c_2$
considered in Ref.~\cite{Huneke2013a}.  Thus, the measurement of further
cumulants will strengthen the evidence for the correct operation of a
steady-state CTAP protocol.

\subsection{Landau-Zener interference}

\begin{figure}
\centerline{\includegraphics[width=.9\columnwidth]{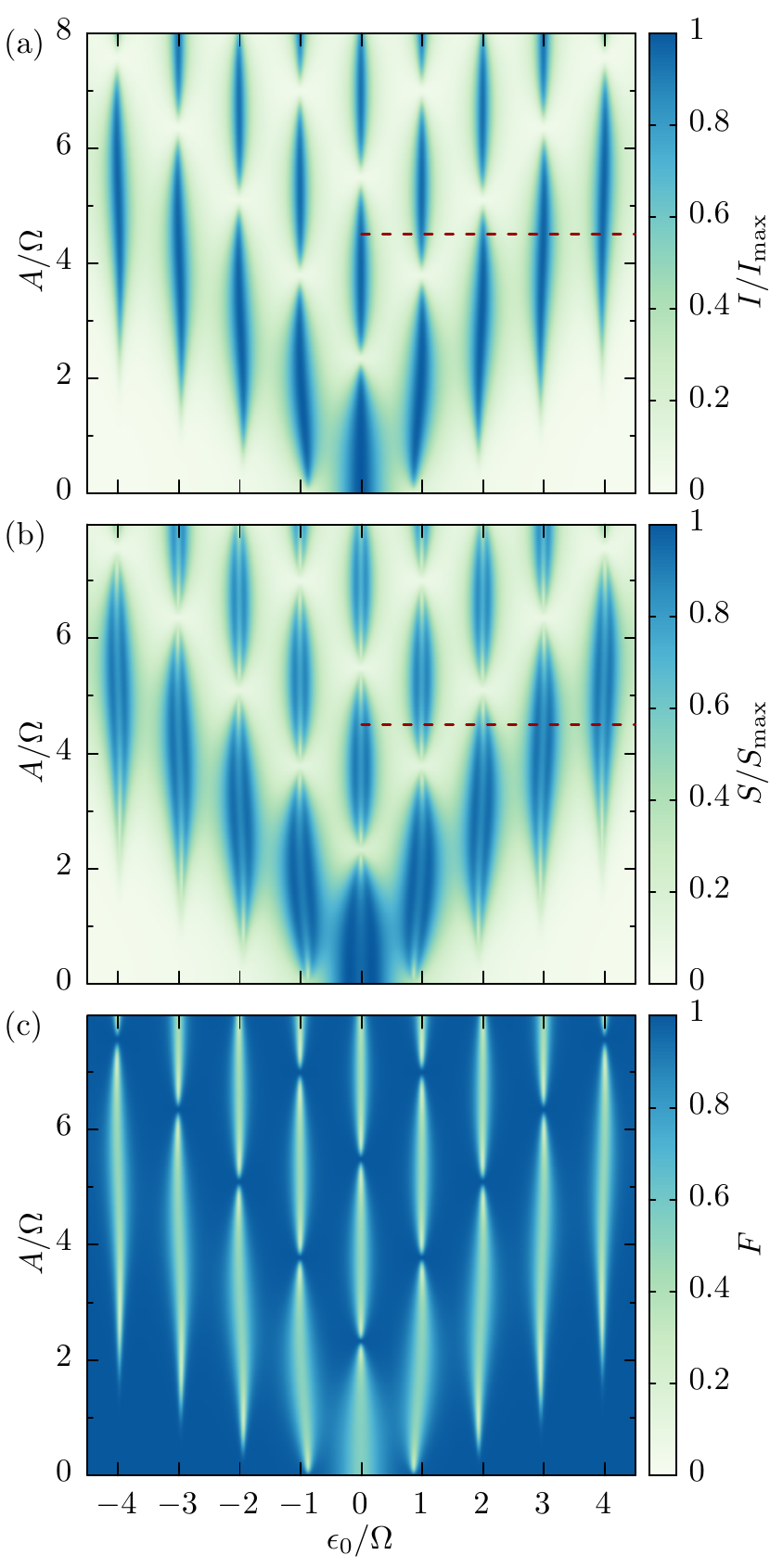}}
\caption{Average current $c_1$ (a), zero-frequency noise $c_2$ (b), and
Fano factor $c_2/c_1$ (c) for a strongly biased driven double quantum dot
as a function of the detuning $\epsilon$ and the driving amplitude $A$.
The driving frequency and the dot-lead tunnel rates are $\Omega=2\Delta$
and $\Gamma_S=\Gamma_D=0.15\Delta$, respectively.  The dashed horizontal
lines mark the amplitude considered in Fig.~\ref{fig:lzsm}.
}
\label{fig:lzsm2D}
\end{figure}
\begin{figure}
\centerline{\includegraphics[width=.9\columnwidth]{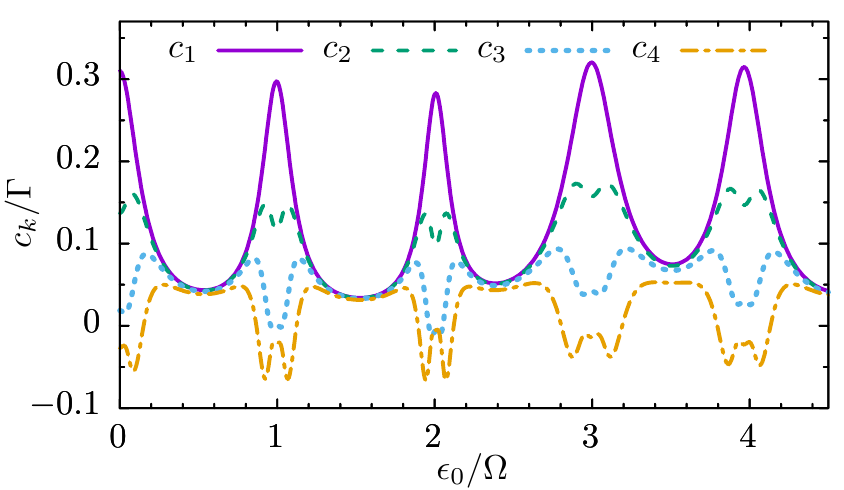}}
\caption{First four cumulants $c_k$ for the LZSM interference
patterns for the driving amplitude $A=4.5\Omega$ marked
in Fig.~\ref{fig:lzsm2D}(a) by a horizontal line.
}
\label{fig:lzsm}
\end{figure}

A paradigmatic example for time-dependent quantum mechanics is
a two-level system with the single-particle Hamiltonian
\begin{equation}
H(t) = \frac{1}{2}
\begin{pmatrix} \epsilon(t) & \Delta \\
\Delta & -\epsilon(t) \end{pmatrix} ,
\label{Hlzsm}
\end{equation}
the tunnel matrix element $\Delta$, and the time-dependent bias
\begin{equation}
\epsilon(t) = \epsilon_0 + A\cos(\Omega t).
\end{equation}
For driving amplitudes $A\gtrsim\epsilon_0$, the eigenenergies of $H(t)$
as a function of time form avoided crossings.  At these crossings, an
electron may perform Landau-Zener transitions, such that repeated sweeps
lead to the so-called LZSM interference.  In a closed system, this is
visible in a characteristic pattern of the population as a function of the
detuning $\epsilon_0$ and the amplitude $A$ \cite{Shevchenko2010a}.  Having
been measured originally for the population of superconducting qubits
\cite{Oliver2005a, Sillanpaa2006a}, such patterns have been found also for
the current in a biased open double quantum dot \cite{Stehlik2012a,
Forster2014a}.  For deeper understanding, we extend previous results for
the average current to a study of current cumulants.

Figure~\ref{fig:lzsm2D}(a) shows the LZSM interference pattern for the
time-averaged current, i.e., the first cumulant $c_1$.  It exhibits the
typical structure found in the high-frequency limit, namely Lorentzian
resonance peaks which by and large are modulated along the $A$-axis by the
squares of Bessel functions \cite{Forster2014a}.  For the second cumulant
[Fig.~\ref{fig:lzsm2D}(b)], the corresponding peaks split into double peaks
whose local minima coincide with the current maxima.  As a consequence,
the corresponding Fano factor [Fig.~\ref{fig:lzsm2D}(c)] assumes clearly
sub Poissonian values of $F_1\approx 1/2$, while off the resonance, the
Fano factor indicates Poissonian transport.

For a closer and more quantitative investigation, we depict in
Fig.~\ref{fig:lzsm} the first 4 cumulants as a function of the detuning
$\epsilon_0$ for constant driving amplitude.  On the one hand, this
highlights the double peak structure of $c_2$ and indicates that at the edge
of the current peaks $c_2\approx c_1$ which corresponds to the Poissonian
$F_1\approx1$.  The third and the fourth cumulants possess a similar double
peak structure, where the magnitude of the $c_k$ diminishes with the order
$k$.  This affirms the low-noise properties of resonantly driven transport
in coupled quantum dots~\cite{Strass2005b}.

\section{Conclusions}
\label{sec:conclusions}

We have developed a method for the iterative computation of current
cumulants for conductors described by a time-dependent Markovian master
equation.  For such transport problems the only generic way to obtain a
solution is a numerical propagation while generally eigenvalue-based
methods are not applicable.  Our scheme is based on a hierarchy of density
operator like objects truncated according to desired number of cumulants.
The cumulants follow in a direct manner by taking the trace.  As compared
to the propagation of a number-resolved density matrix, our scheme
possesses two advantages.  First, it generally gets along with a
significantly smaller set of equations. Second, there is no need to compute
the cumulants from the moments, a numerically critical task that may involve
computing small differences of much larger numbers.

As a test bench, we have employed two time-independent master equations
which can be solved also with previously known eigenvalue-based methods.
This indicated that our scheme provides reliable results for roughly the
first 10 cumulants even for challenging test cases. For less demanding
situations, computing more than 30 cumulants is feasible.  Thus, we reach
orders way beyond the present experimental needs.

We have applied our scheme to two time-dependent systems of recent
interest.  For steady-state CTAP, we have found that not only the second
cumulant, but also higher ones correlate with the population of the middle
dot.  Therefore they may provide additional evidence for the correct
operation of a CTAP protocol.  A similar conclusion can be drawn for
Landau-Zener interference patterns of the current in open double quantum
dots.  The higher-order cumulants substantiate the conclusions drawn from
studies of the Fano factor.

In this spirit, our approach enables studies of the current noise for
time-dependent transport beyond the second cumulant with a moderate
computational effort.  This may provide additional insight to the
underlying transport mechanisms and a deeper understanding of the electron
dynamics controlled by arbitrarily shaped pulses.

\begin{acknowledgments}

This work was supported by the Spanish Ministry of Economy and
Competitiveness via Grant No.\ MAT2014-58241-P and by the DFG via SFB~689.
\end{acknowledgments}


\begin{thebibliography}{37}
\bibitem [{onH()}]{onHuneke}%
  \BibitemOpen
  \href@noop {} {}\bibinfo {note} {Notice that in the figures and captions of
  Ref.~\cite{Huneke2013a}, all times should be multiplied by a factor 20, while
  the rates $\Gamma$ and $\gamma$ should be divided by 10.}\BibitemShut {Stop}%
\end{thebibliography}
%

\end{document}